# Sub-Nyquist time-domain surface-enhanced Raman mapping

Ting Wang[1,*], I. Brian Becerril-Castro[2,*], Ana Sousa-Castillo[3], Miguel A. Correa-Duarte[3,4,5], Ramón A. Alvarez-Puebla[2,6], Matz Liebel[1,a]

[1] Department of Physics and Astronomy, Vrije Universiteit Amsterdam, De Boelelaan 1100, Amsterdam 1081 HZ, The Netherlands

[2] Department of Physical and Inorganic Chemistry, Universitat Rovira i Virgili, Carrer Marcel·lí Domingo 2-4-6, 43007, Tarragona, Spain

[3] CINBIO, Universidade de Vigo, Vigo 36310, Spain

[4] Southern Galicia Institute of Health Research (IISGS), Vigo 36310, Spain

[5] Biomedical Research Networking Center for Mental Health (CIBERSAM) , Vigo 36310, Spain

[6] ICREA, Passeig Lluís Companys 23, 08010, Barcelona, Spain

[*] Both authors contributed equally to this work.

[a] email: m.liebel@vu.nl

## Abstract

Surface-enhanced Raman scattering (SERS) combines analyte-specificity and single-molecule sensitivity, but its potential is limited by slow readout where sophisticated nanosensors are analysed in a serial fashion, one particle at a time. We introduce SERS lock-in sampling to resolve the decades-old trade-off between spectral resolution and widefield imaging. By leveraging the inherent sparsity of Raman spectra, we demonstrate that a simple digital lock-in scheme allows high-quality chemical imaging far beyond the Nyquist-Shannon limit. Our approach integrates an in-situ temporal reference to transform mechanical jitter into an exploitable feature, enabling near-random sampling. We validate SERS lock-in sampling through the multiplexed and simultaneous imaging of thousands of individual SERS-encoded sensors, achieving an orders-of-magnitude throughput-increase over the state-of-the-art. Furthermore, we demonstrate volumetric 3D chemical imaging in biomedically relevant matrices. This robust, computationally simple strategy transforms SERS from a point-observation tool to an imaging modality for clinical diagnostics and real-time chemical observations.

## Introduction,

Surface-enhanced Raman scattering (SERS) spectroscopy is a powerful analytical technique, offering non-destructive, multiplexed measurements with high molecular specificity and single-molecule sensitivity[1,2,3,4]. SERS elegantly merges nanotechnology, analytical chemistry and spectroscopy[4,5,6,7] with widespread applications across disciplines comprising trace analysis[8], environmental sensing and monitoring[9], plant physiology[10], bioimaging[11], diagnostics[12] and catalysis[13,14] as well as intraoperative imaging[15]. Despite 50 years of research and the scalability of colloidal fabrication, commercial adoption of SERS remains limited. This is less due to the sensors themselves than to their slow readout. The field still relies heavily on confocal Raman microscopes, which record spectra through sequential point-by-point raster scanning. This modality restricts acquisition speed, throughput, and field-of-view, making it impractical to interrogate large areas or capture the intrinsic spectral heterogeneity of SERS surfaces in meaningful timescales. Ultimately, this bottleneck hinders the technique's primary potential: rapid, high-throughput imaging and massive spectral multiplexing via tailored sensor libraries.

This throughput limitation is a widely recognised problem, and many strategies have been put forward to address it[14,16,17,18,19,20,21]. Yet, these advances only provide partial solutions. Rapid point-scanning remains constrained by the trade-off between maximising signal and avoiding illumination-induced sample damage[14,17]. Line-scanning increases efficiency but requires complex scanning and dispersed detection hardware[18,19,20]. Hybrid widefield approaches alleviate some limitations but are generally restricted to sparse samples[21]. More fundamentally, the bottleneck lies in detector architecture. Ture spatially resolved spectral imaging would require a three-dimensional camera. While Fourier-transform imaging provides a powerful alternative by encoding spectral data in the time domain[22], its practical utility is limited by slow acquisition speeds. Because Raman scattering is inherently weak, frame rates often drop below 10 Hz. Consequently, acquiring the thousands of frames necessary to satisfy Nyquist sampling and maintain spectral resolution can take tens of minutes. This prolonged duration imposes stringent requirements on positional stability and measurement control[18,22,23,24], offsetting the conceptual advantages of Fourier-transform approaches.

In this work, we present SERS lock-in sampling, a time-domain strategy that yields high-quality spectra from only a few hundred non-uniformly sampled images. This method achieves widefield SERS imaging across large fields-of-view in less than 30 seconds. By reducing both acquisition time and experimental complexity, our approach converts slow, niche Fourier-transform techniques into a practical high-throughput platform. Below, we detail the underlying principles and experimental validation before demonstrating the technique's performance via highly multiplexed 2D and 3D imaging.

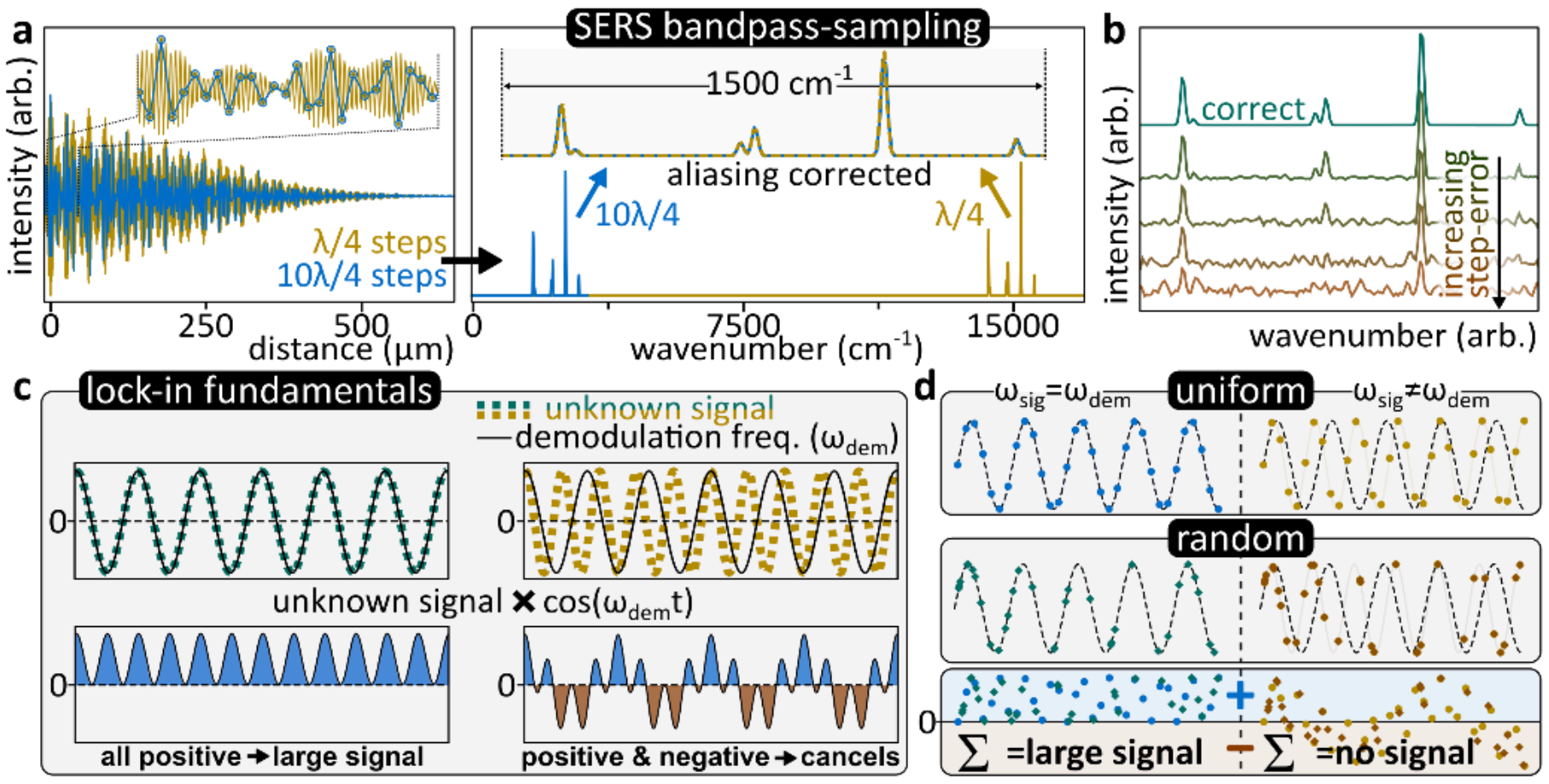

**Figure 1, Concept of SERS lock-in sampling.** a) Simulated interferogram for 6 distinct Raman bands obtained as Stokes spectra with 632.8 nm illumination. Sampling the interferogram at λ/4 (yellow) and 10λ/4 (blue) yields distinct interferograms (left) as well as spectra (right, bottom). When accounting for aliasing-induced changes in wavenumber axis, however, the spectral fingerprint regions contain identical information (right, top). b) Step-size inaccuracies rapidly reduce spectral quality, a major problem of traditional Fourier-transform Raman spectroscopy. c) An unknown signal oscillating at $\omega_{sig}$ (green/yellow dashed line) is multiplied by a periodic reference oscillating at the demodulation frequency $\omega_{dem}$ (black line). The product (red/blue) is integrated over time. When $\omega_{sig} = \omega_{dem}$ (left), the product is positive at all times and accumulates to a large signal. When $\omega_{sig} \neq \omega_{dem}$ (right), positive and negative contributions cancel, yielding near-zero or no signal. d) The same principle applies irrespective of the sampling strategy. Dots and diamonds represent the individual measured data points, sampled either uniformly (top) or randomly (bottom). Temporal integration of the multiplied data yields large signals for $\omega_{sig} = \omega_{dem}$ and near-zero values for $\omega_{sig} \neq \omega_{dem}$.

Lock-in sampling is based on two innovations to: i) circumvent the Nyquist limit by explicitly moving to a bandpass-based strategy and, ii) enable random time-domain sampling. Figure 1a illustrates how spectra can be sampled at rates that nominally violate the Nyquist-Shannon theorem. The key is information-content: Raman spectra are frequency-shifted relative to a Rayleigh frequency thus moving them into the high wavenumber region of visible/near-infrared light. However, the underlying vibrational frequencies are of much lower energy. For a HeNe-laser at 632.8 nm spectra are centred around 15000 $cm^{-1}$ (660 nm) but the spectral range of interest only comprises approximately 1500 $cm^{-1}$ (6600 nm). Sampling the former requires <330 nm steps. The latter could, in principle, be sampled with very large, <3300 nm, steps which explains why Fourier-transform infrared is a popular methodology as absolute vibrational frequencies are measured, not shifts. Figure 1a contrasts Nyquist-limited sampling with a severely undersampled approach. When accounting for aliasing, both approaches yield identical spectral features (Figure 1a, right). These observations demonstrate that bandwidth rather than absolute frequencies dictate sampling requirements, a principle central to bandpass sampling strategies. An important caveat is position-uncertainty, for example due to vibrations or drift, which dramatically impact the spectral quality (Figure 1b). In other words, time-domain Raman measurements require nanometre position accuracy and precision, a delicate task given that millimetre-long scans are necessary.

We address these limitations by introducing SERS lock-in sampling. Inspired by the lock-in amplifier, our method extracts signals by multiplying a time-domain input with an oscillatory reference followed by temporal integration. The underlying working principle of this approach is schematically outlined in Figure 1c, comparing two cases: i) matching signal and reference frequencies and ii) non-matching frequencies. For the former, the product of signal and reference yields only positive values thus resulting in a large integrated output signal. Non-matching frequencies yield positive and negative values and the output integral is near zero (Figure 1c). The traditional lock-in amplifier implements this concept electronically with uniform sampling. We realised that identical results are obtained with random sampling provided the time-points are precisely known (Figure 1d). Mathematically, this serves as a Fourier-transform analogue calculated over discrete, non-uniform intervals. This approach enables high-quality SERS reconstruction from only a few images without stringent positional control. What is needed, however, is accurate and precise knowledge of the exact temporal sampling. Below we introduce a straightforward strategy to access this knowledge.

## Results and Discussion,

Figure 2a outlines the key experimental aspects of lock-in widefield SERS imaging, enabling fast and high-throughput acquisition of thousands of SERS spectra in parallel. In the following, we will briefly describe the function of the system's three core components: i) a conventional widefield microscope; ii) a Michelson interferometer and; iii) a reference channel that measures the pathlength-difference, or time-delay, between the interferometer arms. In the widefield microscope, a narrowband laser illuminates the sample and SERS signals are collected and subsequently imaged (Methods). A Michelson interferometer, placed into the imaging channel, generates two image-copies that interfere on the detector. Acquiring an image sequence while systematically changing the time-delay between the image copies accesses the spectral domain, as outlined in Figure 1. A reference channel based on the co-propagating output of a broadband super luminescent diode and a spectrometer *in-situ* measures the time-delay introduced by the Michelson interferometer (Methods, Supplementary Information S1). In simple terms, the spectrally resolved measurement of the super luminescent diode directly accesses the time-delay via distinct, time-delay dependent, spectral modulation, a direct consequence of time-energy uncertainty[25, 26, 27]. Combined, the three components enable rapid interferogram acquisition without relying on mechanically precise or uniform scanning (Supplementary Information S1).

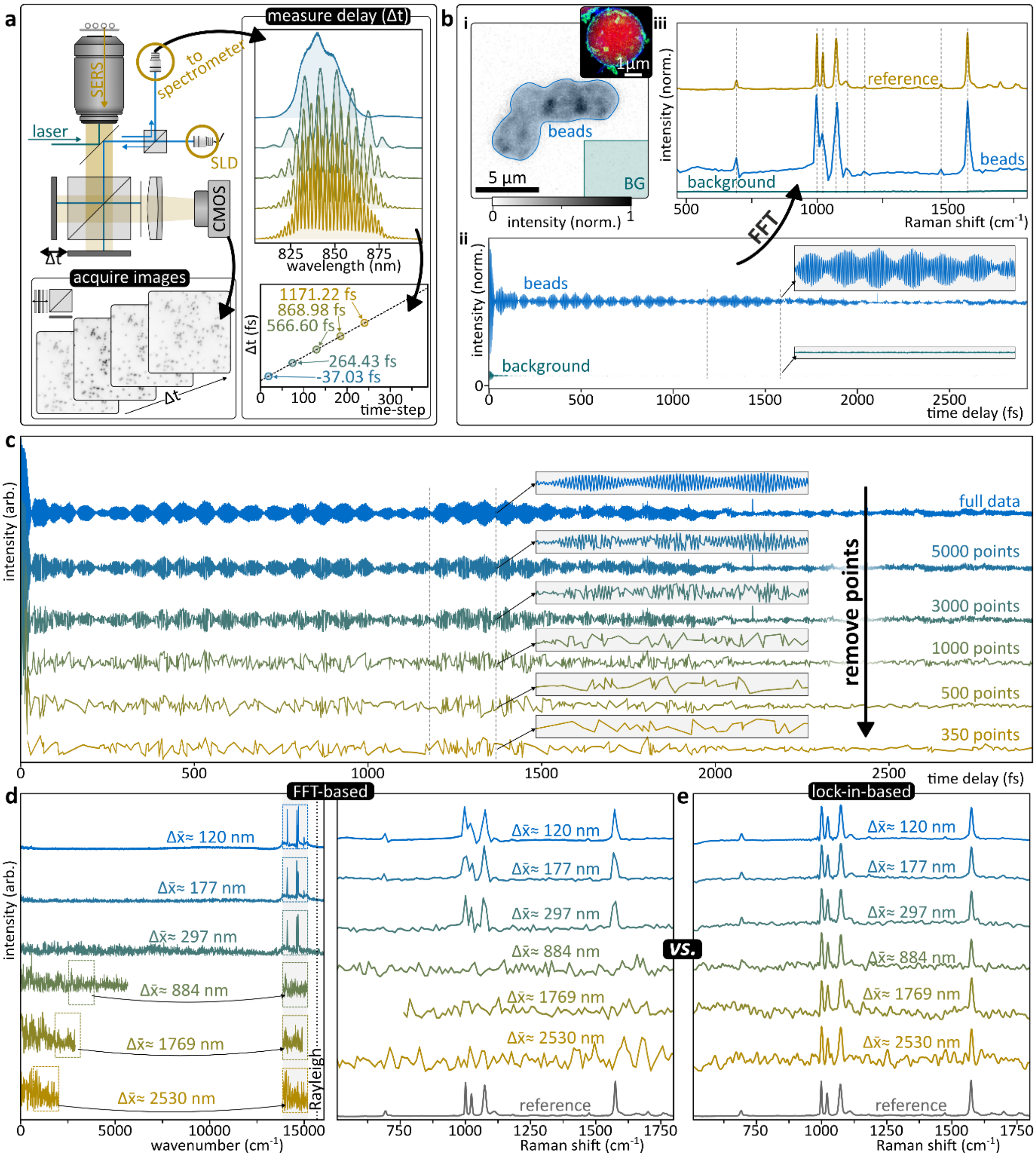

**Figure 2, Experimental implementation and advantages of SERS lock-in sampling.** a) Schematic of the experimental implementation enabled by the combination of a conventional widefield microscope, a Michelson interferometer and a spectral reference measurement based on a co-propagating super luminescent diode (SLD). b) Sample image alongside EDX (Energy-Dispersive X-ray spectroscopy) mapping (Ag: green, C: red, Si: blue) (i) and interferograms (ii) obtained by summing the indicated spatial regions for a conventional Fourier-transform based measurement. Spectra of large 3 µm PS@Ag beads coated with benzenethiol (blue) and a background region (green) are compared to a reference Raman spectrum measured on the as-fabricated bead with a confocal Raman system (yellow, Methods). c) Set of time-traces obtained by randomly removing data points from the data shown in b). d) Spectra obtained via Fast Fourier transformations of c). The absolute wavenumber axis is based on a mean step-size estimate obtained from the maximum time-delay and the number of data points. The box (left, coloured) indicates the region where the Raman spectrum is expected, the zoom in (right) compares the resulting spectra with a reference spectrum, acquired with a confocal Raman system (Methods). e) Spectra obtained from c) via lock-in sampling highlighting robustness even under extreme, totally random sampling conditions. The noise-increase is due to shot noise as nominally fewer photons are detected per spectrum.

To showcase the advantages of SERS lock-in sampling, it is instructive to perform a conventional Fourier-transform experiment. Towards this goal, we selected benzenethiol functionalised SERS-active 3 µm silver covered beads (3 µm PS@Ag-BT, Methods) as a proof-of-concept system. These beads provide bright, stable and spatially well-defined SERS responses to facilitate bead-level identification and spatially resolved analysis without potential issues arising from blinking-instabilities or shot-noise. Figure 2b shows a representative widefield sample-image alongside EDX mapping and spectral interferograms obtained in regions of beads and background. We recorded a time-delay dependent image stack under Nyquist-compliant sampling conditions acquiring 7359 images with 0.12 µm steps. A Fast Fourier transform, assuming constant and perfect stepping, yielded Raman bands at 998 $cm^{-1}$, 1022 $cm^{-1}$, 1072 $cm^{-1}$, and 1573 $cm^{-1}$, in agreement with literature[28], as well as a featureless background spectrum with negligible signal.

To highlight the advantages of SERS lock-in sampling, we removed randomly selected time-points from the long interferogram thus simulating increasingly coarse and completely arbitrary stepping (Figure 2c), an extremely challenging sampling scenario compared to mere undersampling on a regular grid. Importantly, we retained the associated temporal positions as retrieved by the reference measurement. At the extremes, only 350 steps covered almost 3 ps and the time-traces started resembling random noise. As expected, Fourier transform based spectral recovery struggled with the nonuniformly sampled data, even when accounting for the shifts in absolute wavenumbers due to aliasing (Figure 2d, Supplementary information S2). Importantly, even under Nyquist-compliant conditions the spectra were of poor quality, likely due to time-delay jitter (Figure 1b, 2d)[24]. In stark contrast, SERS lock-in retrieval recovered spectra of excellent quality from the interferograms, irrespective of the sampling conditions (Figure 2e, Methods, Supplementary Information 5). Even randomly selecting 350 time points with a mean delay that is approximately 10-fold beyond the Nyquist limit yielded spectra that closely resembled the ground-truth measurements with the noise increase being a direct result of the reduction in total photons collected.

Figure 2 demonstrates that lock-in sampling retrieves high-quality spectra under extreme sampling conditions. However, realistic experiments are likely to provide somewhat uniform stepping with a controlled level of uncertainty and jitter. We thus performed simulations balancing spectral resolution, spectral observation window and the number of time-points to determine optimum conditions (Supplementary Information S3) and identified 350 sampling points with a mean step size of 2.6 µm as a sweet-spot. These settings allow measuring high-quality spectra in the bandpass-determined 465-1975 $cm^{-1}$ (Methods) spectral range with a spectral resolution of approximately 10 $cm^{-1}$, as governed by the 910 µm delay covered by the interferometer assuming Gaussian line shapes, in a short time. Typical integration times of 50-200 ms per image ensure operation above the dark noise of even cheap industrial CMOS cameras and translate to total acquisition times for a full spectrally resolved SERS image of around 30 seconds to 1 minute.

Thus far, we presented proof-of-concept and validation experiments using a single cluster of SERS-active 3 µm PS@Ag-BT beads, but the true strength of SERS lock-in sampling is its widefield nature that allows simultaneously interrogating very large fields-of-view. To validate these capabilities, we prepared samples composed of 1 µm $SiO_2$@Ag microbeads encoded with three distinct Raman reporters: 4-mercaptobenzoic acid (4-MBA), 4-methylbenzenethiol (4-MBT), or 3-fluorothiophenol (3-FTP) (Methods). These molecules were selected for their strong Ag-thiol interactions and distinct, intense SERS signatures. Representative SEM images of the beads, alongside reference and lock-in imaging-based Raman spectra are shown in Figure 3a. We then dispersed all three SERS reporters on the same substrate and performed widefield measurements over an extreme field-of-view measuring approximately 390×390 $µm^2$ (Methods) to provide proof-of-concept validation for spectrally and spatially multiplexed applications. For the relatively sparse sample shown in Figure 3b, we were able to simultaneously measure the spectra of >2300 microbeads and then classify them based on their spectral signatures (Figure 3c). Importantly, even under these extreme conditions we achieved sufficient signal-to-noise ratios for classification. To put the multiplexing capabilities into context, the measurement reported in Figure 3a-c required 70 seconds, a conventional confocal Raman microscope would take approximately 16 hours to raster-scan the same area with an insufficiently large step size of 0.5 μm and an integration time of 100 ms per point.

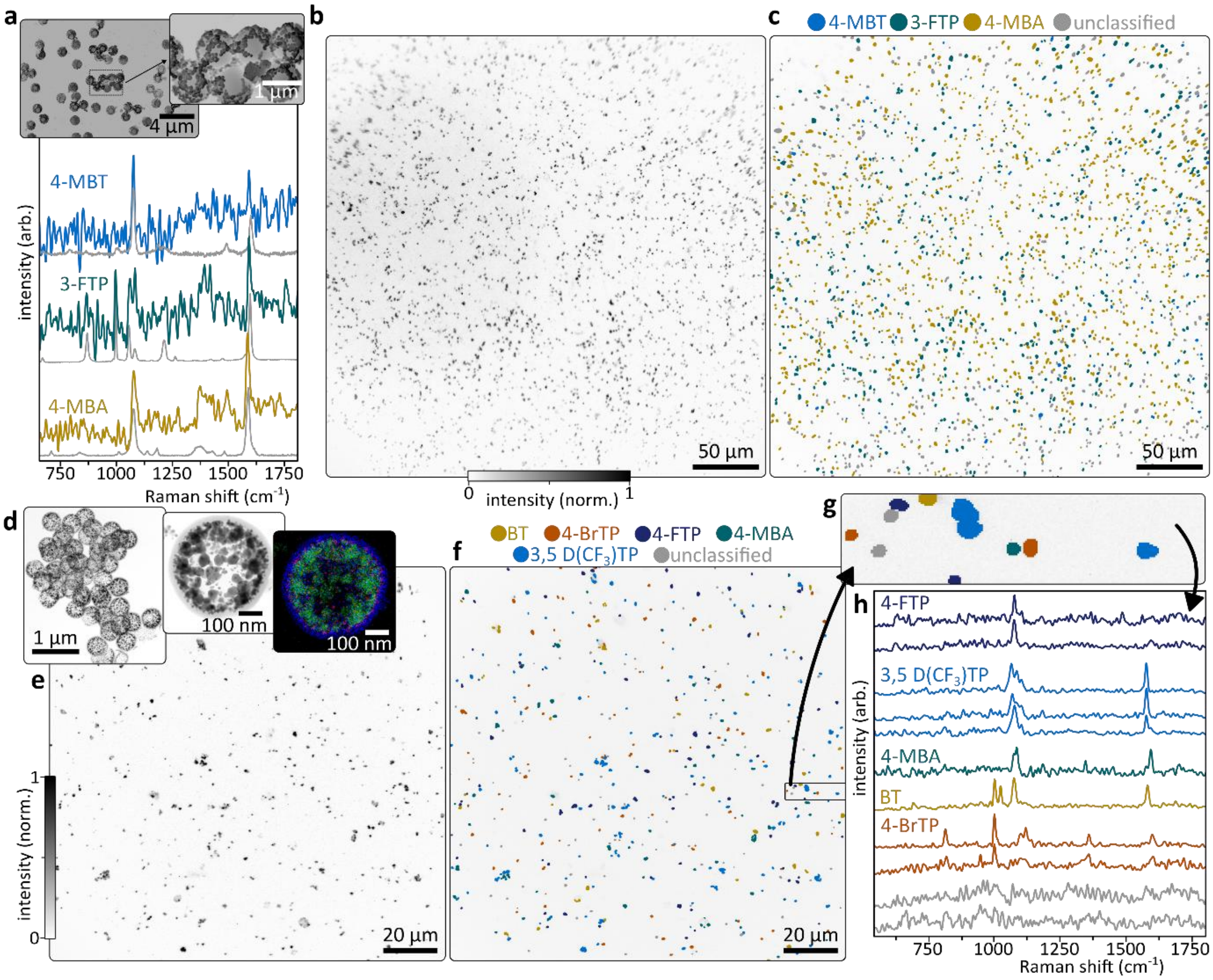

**Figure 3, Extreme multiplexing of multiple analyte systems.** a) Scanning electron microscopy (SEM) image and magnification of 1 μm $SiO_2$@Ag beads alongside representative conventional (grey) and lock-in imaging-based Raman spectra of the three Raman codes. b) Widefield intensity image of a mixed-analyte sample containing 4-MBA, 3-FTP and 4-MBT beads. c) Classification of the individual particles shown in b) based on their spectral signatures. d) High Resolution Transmission electron microscopy (HRTEM) image and magnification of the SERS-encoded 400 nm Ag nanocapsules (AgNCaps) alongside EDX maps of AgNCaps for Ag, Au and Si. e) Widefield intensity image of a mixed-analyte sample containing beads encoded with 4-FTP, 4-BrTP, 3,5 D($CF_3$)TP, BT and 4-MBA. f) Corresponding spectral classification and g) magnification of a representative region. h) Normalised SERS spectra of the SERS-codes highlighted in g).

Yet, these results do not reflect the ultimate performance of our platform, as the 1 μm targets remain far above the diffraction limit. To extend our detection capability toward the diffraction limit of the microscope, we prepared ~400 nm Ag nanocapsules (AgNCaps, Figure 3d). EDX mapping analysis of the AgNCaps (Figure 3d, inset) confirmed the presence and distribution of nanostructured Ag around the Au seeds within the silica framework. Having verified the composition of these highly active SERS platforms, AgNCaps were functionalized with five thiolated Raman reporters (Methods). Samples containing all five SERS codes were then interrogated over a 175×175 $μm^2$ field-of-view and classified based on their spectral signatures as in the previous example (Figure 3e, f, Methods). Albeit the presence of small aggregates, individual capsules remained resolvable (Figure 3g), showing high-quality spectra with readily identifiable SERS bands (Figure 3h) as a direct result of the nominal ~6-fold increase in illumination fluence compared to the data presented in Figures 3b, c.

Motivated by the results on flat, quasi-2D, samples we decided to make first attempts towards complex 3D structures, as enabled by the fast spectral image acquisition. To this end, we suspended 1 µm $SiO_2$@Ag microbeads coded with four different Raman reporters in a 4% agarose gel and acquired a 3D image stack spanning 150.7×150.7×50 $\mu m^3$ (Methods). Figure 4a shows representative images of the image stack. Figure 4b depicts an in-depth analysis of a single z-plane, following the same strategy as previously employed in the 2D case (Figure 3) but taking z-defocusing explicitly into account (Methods, Supplementary Information S4). We were able to classify individual SERS codes based on their distinct fingerprints. Interestingly, we observed background signals reminiscent of a mixture of the four SERS-codes, an aspect we attribute to out-of-focus contributions from all codes contained in the extended agarose gel (Figure 4b). Moving from a single plane, we now focus our attention on the full 3D data-stack where 3D localisation and classification recover a volumetric distribution of the suspended particles (Figure 4c, Methods). From a single stack, comprising 51 images we recovered 2199 particles which appeared uniformly distributed within the hydrogel. A closer look at the yz-projection (Figure 4c) revealed increased detection-probability towards the bottom of the sample, an aspect that could be either due to increased aberrations inside the hydrogel, thus lowering the detection probability, or due to gravity-induced particle sedimentation during hydrogel formation.

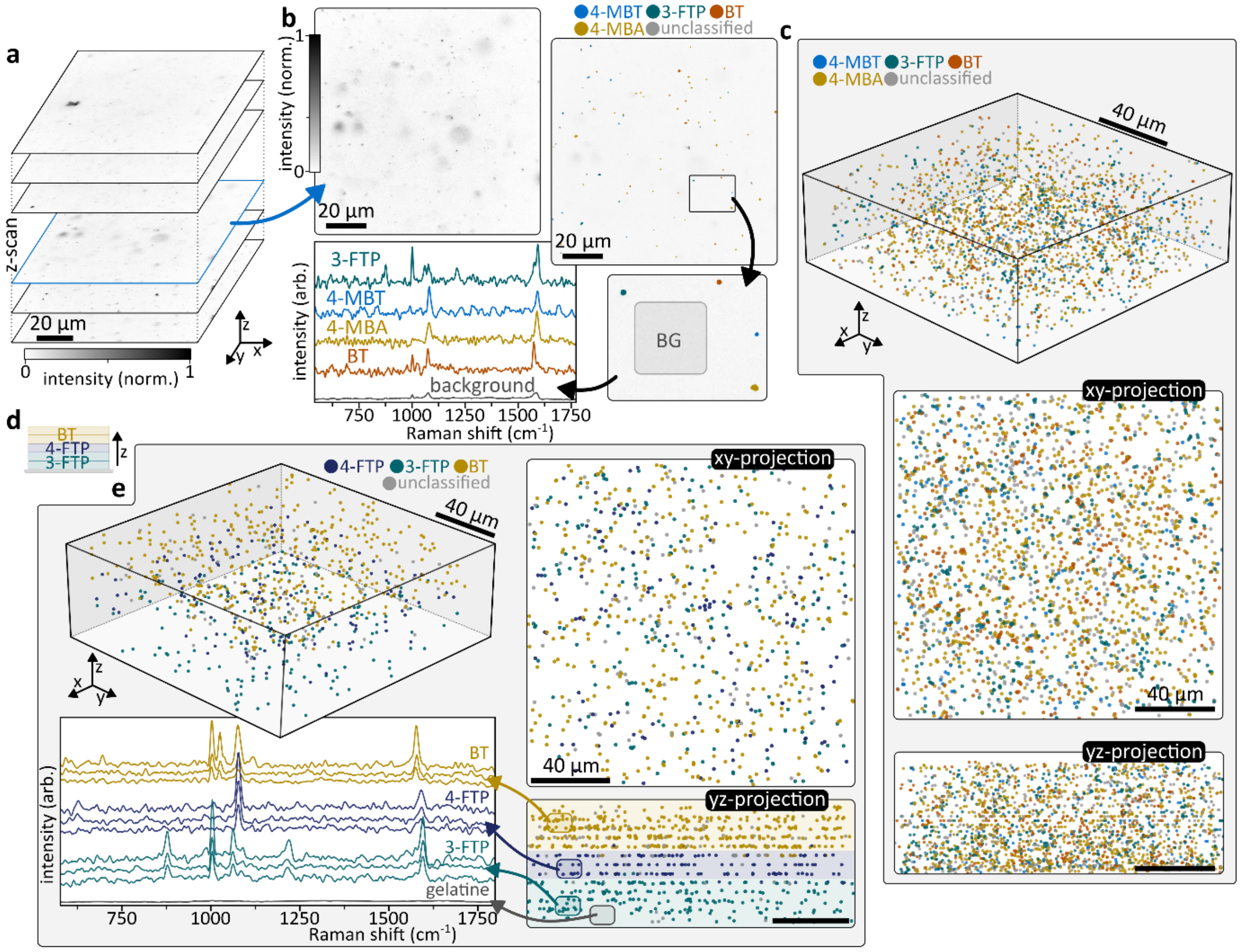

**Figure 4, Multiplexing and 3D sampling of complex suspensions.** a) Schematic illustration of the 3D scanning procedure. b) Widefield image of a representative z-plane within the 3D agarose sample containing 4-MBA, 3-

FTP, 4-MBT and BT microbeads alongside classified image and representative single-particle SERS spectra. c) 3D volumetric representation of the classification results alongside xy- and yz-projects. A total of 51 planes were used. d) Schematic of the spatial organisation of the layered gelatine sample. e) 3D classification alongside projections and representative spectra, as b, c), based on many individual nanoaggregate observations within the gelatine sample.

The 3D agarose measurements highlight our ability to interrogate sample densities and geometries that are essentially impossible to address by other means. However, agarose samples have little diagnostic or clinical relevance and localising randomly distributed mixtures of particles does not demonstrate our ability to 3D localise structural arrangements correctly. To address these shortcomings, we prepared SERS code suspensions in precisely defined, and stacked, gelatine layers as shown in Figure 4d (Methods). A 3D representation of the classified 3D image-stack alongside xy- and yz-projections revealed that we were indeed capable of spatially distinguishing the as-fabricated layer structure (Figure 4e). In the xy-plane we observed a near random particle distribution, in stark contrast to the spatially defined 3-FTP/4-FTP/BT arrangement in the yz-plane. A closer look at the spectra of individual particles showed satisfactory distinction between the three SERS-codes and far-superior signal compared to the weakly Raman scattering gelatine matrix.

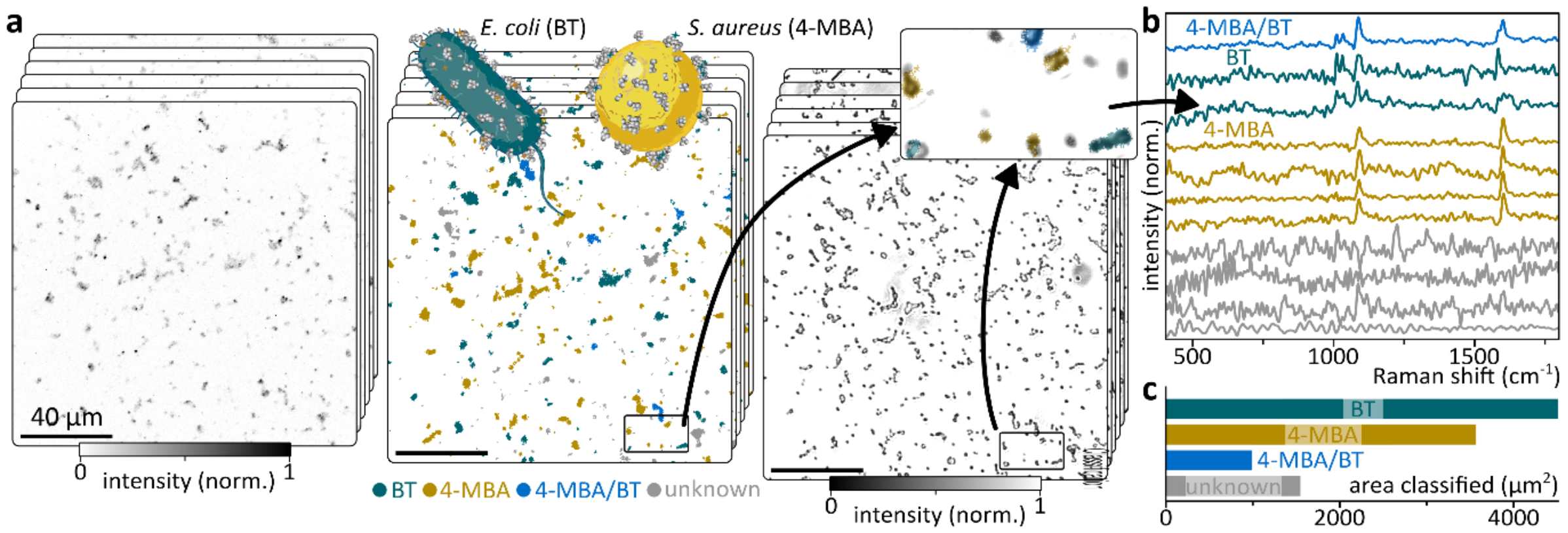

**Figure 5, Practical validation of multiplexed bacterial SERS classification**. a) Schematic illustration of the labelled bacteria@AgNP substrates, with *S. aureus* labelled by 4-MBA and *E. coli* labelled by BT. SERS (left), classification (centre) and brightfield images (right) are shown including an inset overlaying classification and brightfield images (inset). b) Normalized SERS spectra of the encoded bacteria. c) Comparison of the total classified area for *S. aureus* (4-MBA), *E. coli* (BT), co-localized *S. aureus* and *E. coli* (4-MBA/BT), and unclassified regions.

Building on the performance demonstrated in 2D and 3D measurements, we evaluated the system using a prepared mixture of SERS-encoded bacteria. For this demonstration, we selected a reporter-based strategy rather than intrinsic bacterial fingerprinting[29]. Our previous work used SERS-encoded Ag nanoparticles functionalized with bacteria-selective antibodies for multiplex detection in biofluids[30, 31]. Here, to evaluate the imaging readout independently of antibody-mediated recognition, *E. coli* and *S. aureus* were separately coated with pre-synthesized Ag nanoparticles and encoded with BT and 4-MBA[32], respectively. The encoded populations were subsequently mixed and deposited onto a coverglass. This experiment evaluates spatially resolved readout of pre-encoded bacterial populations rather than selective identification of unknown bacteria. Figure 5a compares widefield SERS and brightfield images acquired over a 150 × 150 $\mu m^2$ field of view. The brightfield images provide morphological and spatial context of the *E. coli* and *S. aureus*, complementary insight to the SERS images where, the apparent bacterial morphology could be slightly altered as the Ag NPs are somewhat non-uniformly aggregated on the bacterial surfaces. Widefield SERS, however, allows spectral discrimination of the two bacterial species based on their respective SERS labels (Figure 5b). Even though the non-uniform Ag NPs coverage resulted in irregular bacterial shape, following the same spectral classification strategy used for the 2D measurements, the bacterial species could be distinguished using the characteristic SERS signatures of their attached encoded particles. A magnified representative region is shown in Figure 5a, where the classification map is overlayed onto its corresponding brightfield image. Although some bacteria were insufficiently coated with Ag NPs and therefore produced weak or undetectable SERS signals, bacteria with detectable characteristic spectral features could still be reliably classified (Figure 5b). To further assess the robustness of bacterial classification, six different fields of view with a total acquisition time of 3.5 min corresponding to an imaged area of 135,000 $\mu m^2$. Across the six measurements, the total area used for object-classification, based on sufficient signal levels, was 10,611 $\mu m^2$, out of which 3,568 $\mu m^2$ was assigned to *E. coli*, 4,513 $\mu m^2$ to *S. aureus*, and 986 $\mu m^2$ to regions containing overlapping or adjacent signals from both bacterial species (Figure 5c). These results are in qualitatively good agreement with the underlying bacteria concentrations, prior to labelling, where stock solutions of comparable optical densities were used and, further, provide interesting insight regarding throughput increases. For the latter, only sampling the classifiable area with a confocal system at 100 ms integration time and a 1x1 $\mu m^2$ observation area would require an at least 5-fold increase in observation time. Systematically mapping the whole area would require several hours thus further underlying the advantage of SERS lock-in sampling.

## Summary,

A central challenge in spectrally resolved imaging is combining high spectral resolution, large field-of-view and imaging speed. Fourier-transform approaches are often championed as promising candidates, yet in practice they remain constrained by the limited speed of 2D detectors and by the photon budgets required for shot-noise-limited operation. When combined with Nyquist-compliant sampling, these constraints readily push Raman hyperspectral acquisition times into the range of tens of minutes. SERS lock-in sampling overcomes this long-standing bottleneck by enabling rapid widefield acquisition without sacrificing spectral fidelity.

The method exploits a defining property of Raman spectra: the spectral window of interest is narrow relative to the absolute optical frequency. This makes it possible to recover high-quality spectra from only a few hundred images. Although such strategies could in principle be implemented within conventional Fourier-transform frameworks, in practice they are highly susceptible to position uncertainty and jitter, both of which rapidly degrade spectral quality. Non-uniform Fourier-transform approaches may offer an alternative, but in our hands established methods did not provide robust recovery for heavily undersampled data, likely because they still rely on interpolation assumptions linked to near-Nyquist sampling[33]. By contrast, our lock-in retrieval strategy (Figure 1c) enables direct and robust spectral reconstruction under strongly undersampled conditions and therefore constitutes a central technical advance of this work. The second is an experimentally straightforward yet highly precise time-delay measurement scheme that enables accurate lock-in retrieval (Figure 2a). Our methodology is, at its core, based on whitelight interferometry which enables absolute position estimates but remark that alternative approaches based on different reference-measurements can also be combined with lock-in sampling[34, 35, 36]. Together, these advances permit near-random sampling (Figure 2c, e), preserve spectral quality, and enable substantially relaxed hardware requirements.

Experimental validation on both 2D and 3D samples highlights the strong potential of this platform for highly multiplexed SERS imaging. We classified more than 2300 particles across a 390×390 µm field-of-view in approximately 70 s, whereas a conventional confocal Raman system would require an estimated 16 h to raster-scan the same area under comparable conditions. In principle, this area could contain more than 200,000 diffraction-limited SERS particles, pointing to substantial headroom for even higher throughput. These capabilities directly unlock applications where statistically significant observations are highly relevant, such as the identification of microplastics in the context of real-world applications[37, 38]. A direct comparison with state-of-the-art Fourier-transform approaches, notably our previous holographic SERS implementation[24], further underscores the scale of the advance achieved here, with an estimated throughput increase exceeding two orders of magnitude when field-of-view and illumination fluence are taken into account. Finally, the 3D localisation of SERS-encoded particles within structured gelatine layers demonstrates that the method extends beyond planar imaging and opens a practical route towards volumetric chemical imaging in complex samples such as those from medical or environmental origin.

## Conclusion,

In summary, we have implemented and validated SERS lock-in sampling as a robust, high-throughput methodology for spectrally resolved widefield imaging. We believe that this methodology provides highly a promising foundation for imaging based vibrational spectroscopy. To the best of our knowledge, it overcomes the main drawbacks previously associated with Fourier transform based imaging and substantially enhances performance, thus achieving a true optimization rather than merely adapting the technology to a specific niche application. In other words, SERS lock-in sampling solves long standing issues in the field and, as such, presents a widely applicable, useful and robust sampling strategy. Experimentally, implementing SERS lock-in sampling is trivial, as it relies on merely co-propagating a broadband reference beam, derived as a collimated output of a single mode fibre, with the signal of interest, followed by detecting the self-interfering light with a spectrometer.

Given its scalability and ease of implementation, SERS lock-in sampling has the potential to transform SERS from a specialized analytical niche into a practical, high-throughput tool for complex diagnostic and 3D imaging applications. More specifically, SERS lock-in sampling unlocks key-capabilities that have thus far been difficult to impossible to achieve. Examples are broad spatial mapping of analyte distributions to, for example, visualise catalytic processes near interfaces, study tug-of-war chemical interactions between microorganisms, diffusion and transport phenomena in biological samples or rapidly assess tumour margins during surgery[39]. Another application is digital SERS where rapidly assessing thousands of different spatial locations directly impacts detection sensitivity and measurement uncertainties thus further pushing this exciting technology[40, 41]. Finally, we envision application in the context of digital sensor toolkits where tailored mixtures of analyte-specific colloidal probes can be incubated with a sample of interest followed by rapid widefield readout, potentially with a low-cost or even hand-held SERS lock-in reader. While SERS lock-in sampling addresses several long-standing limitations of SERS imaging, several practical considerations remain. More specifically, zero-delay position estimation requires thorough data processing, and the sampling parameters must be carefully selected according to the desired spectral window to avoid aliasing. From a general imaging point of view, weak out-of-focus signals may contribute residual background in 3D imaging, analogous to all other non-confocal implementations. Light-sheet implementations are the logical route towards mitigating these issues. Ultimately, specifically tailored SERS substrates remain key for high-quality analytical applications where the time-domain nature of SERS lock-in sampling offers promising advantages in terms of luminescent-background removal through time-domain filtering at short time delays.

Beyond SERS, our measurement methodology holds great promise for conventional Raman spectroscopy and imaging as well as the closely-related Brillouin microscopies where innovative time-domain strategies are beginning to emerge[42]. Due to its technological simplicity, we see direct applications in the context of turn-key instrumentation which currently relies on complex and costly confocal systems with high-performance spectrometers. Furthermore, emerging Raman methodologies, such as spatially offset Raman spectroscopy that is increasingly being explored in the context of diagnostics and conservational sciences would dramatically benefit from lock-in sampling. Rather than relying on complex excitation and detection geometries[43] lock-in sampled widefield observations directly capture the full information content in a single acquisition. Finally, a direct extension to any phase-locked hyperspectral methodology[44, 45] is within reach where the methodology implemented here is primed to rapidly replace complex and generally extremely costly, hardware-based pathlength stabilisation with a simple reference measurement combined with computational lock-in retrieval.

## Methods,

### Lock-in widefield Raman microscope; general imaging system,

The output of a 638 nm laser (*Integrated Optics*, *638L-41A*) was free-space coupled into a multimode fibre and the fibre-output subsequently imaged into the sample plane of a widefield microscope equipped with a numerical aperture 0.75, 40x objective (*Nikon, N40X-PF*). A 640 nm longpass filter (*AHF Analysentechnik AG, F48-641*) separated the incoupled light from Raman scattering. A 628/2.4 nm bandpass filter (*AHF Analysentechnik AG, F94-639*), placed into the fibre output rejected residual laser light. A combination of a 647 nm longpass (*AHF Analysentechnik AG, F76-649*) and a 685/80 nm bandpass (*AHF Analysentechnik AG, F47-688*) isolated the Raman scattering in the spectral region of interest. An imaging system composed of a 1:1 relay system (200 mm focal length achromatic doublets) followed by a focal length 80 mm (achromatic doublet) formed an image of the sample plane on the detector (*Basler AG, a2A2840-67g5mBAS*) at a nominal magnification of 16x corresponding to 171.2 nm/px. A Michelson interferometer placed between the 1:1 relay and the 80 mm imaging lens generated the image-copies for SERS lock-in sampling. A linear piezo stage (*Precibeo, GO Stage LLS4545*) controlled the pathlength-difference between the two arms of the interferometer. Custom software (python) handled all aspects related to instrument control and image acquisition.

### Lock-in widefield Raman microscope; reference channel,

The collimated output of a fibre coupled super luminescent diode (*Superlum Diodes Ltd, SLD-BLL-S840.35.12S*) first passed a polarizing beam splitter followed by a quarter wave plate to allow efficient in- and out-coupling into the Michelson interferometer by providing a full linear polarisation rotation on the back path. An 800 nm longpass filter (*Thorlabs, FELH0800*) was placed into the beam path prior-to incoupling the light into the imaging channel via an 805 nm longpass dichroic mirror (*Thorlabs, DMLP805*). Upon passing the interferometer, the back-reflected light exited via the polarizing beamsplitter and was coupled into a single mode fibre towards the custom high-resolution spectrometer comprised of a 150 mm achromatic lens, a 600 grooves/mm grating and an 80 mm achromatic lens (all *Thorlabs*). A CMOS camera served as the detector (*Basler, 2-a2A3840-45umBAS*). The spectral interferograms recorded during interferometer operation were then used to extract the time-delay, or optical path-length difference, as outlined in detail in Supplementary Information 1.

### Lock-in spectral recovery,

The spectral intensity, $I(\omega)$, at an arbitrary frequency of interest, $\omega_{int}$, is obtained as:

$$I(\omega) = \max_{\varphi\in[0,2\pi]} \sum_t x(t)\cos[2\pi\omega_{int}t + \varphi]$$

Where *x(t)* is the experimentally measured data, *t* the retrieved time delays and $\varphi$ the phase. For retrieving real/imaginary parts the phase is kept constant at $\varphi = 0, {}^{\pi}/_{2}$. Means of improving the spectral quality by enforcing symmetry around time-zero, which requires precise knowledge of the temporal sampling axis, or phase[46], and accounting for image-position dependent time-delays are discussed in Supplementary Information 5.

**Electronic microscopy characterization,**

EDX maps of AgNCapsules and 3 μm PS@Ag beads, as well as HRTEM images were obtained using a JEOL F200. SEM images of 1μm $SiO_2$@Ag microbeads were obtained with a FEI Verios 460.

**Reference spectra measurement,**

All the reference spectra of SERS sensors were obtained by confocal Raman microscope (InVia, Renishaw, UK) with 632.8 nm laser using the respective SERS particles discussed in the corresponding figures.

**Chemicals and consumables,**

Silver nitrate ($AgNO_3$, 99.9%), L (+)-Ascorbic acid (AA, ⩾99%), Tri-sodium citrate dihydrate (Sodium citrate, ⩾99%), 4-mercaptobenzoic acid (4-MBA, 90%), 3-Fluorothiophenol (3-FTP, 98%), Thiophenol (BT, 99%), 4-methylbenzenethiol (4-MBT, 98%), Hexadecyltrimethylammonium bromide (CTAB, ⩾99%) and low melting point agarose were purchased from Thermo scientific. Poly(allylamine hydrochloride) (PAH, Mw 17500), sodium chloride (NaCl, 99%), Poly-L-Lysine Solution (0.1mg/mL) and 11-Mercaptoundecanoic acid (MUA, 95%) were obtained from Sigma-Aldrich. 3, 5 Bis(trifluoromethyl)thiophenol (3,5 D($CF_3$)TP, 98%), 4-Fluorothiophenol (4-FTP, 97%) were obtained from Alfa Aesar. Ethanol (95%), Sodium hydroxide (NaOH, ⩾97%) and Aqueous ammonia (28 wt%) from Fisher. Tetraethyl orthosilicate (TEOS, 98%) and tetrakis(hydroxymethy1)phosphonium chloride (THPC, 80%) were purchased from Acros Organics. Gelatine (iggy's) was purchased at local supermarket. $SiO_2$ beads solution (5 wt%) was obtained from Microparticles GmbH, Polystyrene (PS) beads (3 μm and 400 nm) where purchased from Ikerlat Polymers. microparticles. All reagents were stored and used as received. 200 μm thickness PDMS film was obtained from Hangzhou weisichuang technology company. MilliQ water (18 mΩ $cm^{-1}$) was used in all aqueous solutions, and all the glassware was cleaned with aqua regia before the experiments. 1.5H cover glasses were cleaned with acetone and ultrapure water prior-to being used in all experiments.

**Silver covered beads (3 μm PS@Ag)**

10 mg of 3 μm PS beads were functionalized using PAH (Mw = 17500). A PAH solution (1 mg/ml) was prepared in a 0.5 M NaCl solution and sonicated during 15 min. 1 mL of PS beads were added to 1 mL of the PAH solution and left under agitation during 30 min. Cleaning of excess PAH was performed with 3 centrifugation steps (800 rpm, 10 min) and final redispersion in 1 mL of water. To achieve complete coverage of the positively charged PS beads, the amount of pre-synthetized Ag NPs (~35 nm)[13] required to fully cover the PS beads (~3 μm) was calculated and multiplied by 4 to ensure an excess of AgNPs. The calculated volume of PS beads was added to the AgNP solution under sonication. After addition, the mixture was thoroughly mixed and left on a shaker overnight. The following day, the PS@Ag were subjected to 3 centrifugation cycles at 800 rpm for 5 minutes and redispersed in Mili-Q water to a final bead concentration of 0.5 mg/mL. A silica shell was achieved through a modified Stöber method following the described in the Nanocapsules (NCaps) section. All the Raman reporters were dissolved in ethanol at a stock concentration of 10 mM.10 μL of 1mM Raman reporter (BT) was added into 1 mL AgNCaps to form SERS amplifiers. Samples were fabricated by drying 10 uL of the PS@Ag suspension on a cleaned cover glass.

**Multiplexed 1 μm $SiO_2$@Ag microbeads on PDMS**

The cleaned cover glasses were treated with plasma cleaner (*Diener electronic GmbH + Co. KG, Zepto Model 4*) to generate negatively charged surfaces. Aminated $SiO_2$ beads (0.2 mg/mL) were deposited onto the plasma-cleaned cover glasses for 2 min, after which unbound beads were removed by gentle washing. The surface density of $SiO_2$ beads was controlled by adjusting the deposition time. The bead-decorated cover glasses were incubated with freshly prepared 80 nm citrated Ag NPs for 24 hours and subsequently washed. 80 nm Ag NPs were synthesized using classic Turkevich method[47] with slight modification. In brief, 100 mL of 1 mM $AgNO_3$ solution were heated to boiling, and 1.2 mL of 1% sodium citrate and 1 mL of 1% ascorbic acid were rapidly added under stirring (800 rpm) for 1 hour. After the reaction mixture was cooled to room temperature. The Ag-coated $SiO_2$ microbeads on the cover glasses were incubated with 1 mM Raman reporter solution for 6 hours. After incubation, each cover glass was washed with ultrapure water and dried. A thin adhesive PDMS film was used to sequentially lift off $SiO_2$@Ag beads functionalized with three different Raman reporter molecules (4-MBA, 3-FTP and 4-MBT) from separate cover glasses, thereby assembling a multiplexed sample containing distinct bead populations on a single PDMS substrate.

**Ag Nanocapsules (AgNCaps)**

As an initial step, 12.5 mg of PS beads (400 nm) were functionalized using PAH following a previous report[48]. In this procedure, PAH was dissolved in NaCl (0.5 M, pH 5.0) to obtain a polymer concentration of 1 mg $mL^{-1}$. A positively charged PAH solution (25 mL) was then added to the PS beads, and the mixture was stirred at room temperature for 30 min. Excess reagents were removed through four cycles of centrifugation and redispersion in water (9000 rpm, 40 min per cycle). This treatment imparted the PS particles with the necessary electrostatic charge for the subsequent adsorption of Au nanoparticles (AuNPs).

AuNPs (~3 nm), were synthesized and subsequently adsorbed following a modified procedure based on the method reported by Duff *et al*[49]. Briefly, 9.1 mL of water was placed in a 20 mL vial under stirring, followed by the sequential addition of NaOH (0.3 mL, 0.2 M), THPC (0.2 mL, 84.45 mM), and $HAuCl_4$ solution (0.4 mL, 25 mM). Upon addition of the gold precursor, the solution rapidly turned dark brown, indicating nanoparticle formation. The mixture was stirred for approximately 15 min, and the resulting colloidal suspension was stored at 4 °C until further use. A dispersion of AuNPs (10.0 mL, [Au] = $1 \times 10^{-3}$ M) was slowly introduced dropwise into the PS bead suspension (50 mL, 0.25 mg $mL^{-1}$) under sonication. The mixture was then stirred at room temperature (400 rpm) for 30 min to promote nanoparticle adsorption onto the PS surface. Excess AuNPs were subsequently removed by three successive cycles of centrifugation and redispersion in Milli-Q water (3500 rpm, 40 min per cycle). After the final washing step, the collected sediment was redispersed in water (12.5 mL), yielding a suspension with a final concentration of 1.00 mg $mL^{-1}$ relative to the PS template. The silica shell was deposited following a previously reported method.[48] Briefly, the PS@Au dispersion (12.5 mL) was slowly introduced dropwise under sonication into a mixture containing CTAB (312.5 mg), Milli-Q water (125 mL), ethanol (50 mL), and $NH_4OH$ (1.5 mL,28 wt%). After 15 min of sonication, an ethanolic solution of TEOS (2.06 mL, 5% v/v,) was added dropwise while maintaining sonication. The reaction mixture was subsequently stirred for 24 h to allow uniform growth of the silica layer. The resulting PS@Au@$SiO_2$ particles were purified by four cycles of centrifugation and redispersion in ethanol (3500 rpm, 20 min per cycle). Finally, the organic components (PS and CTAB) were removed by calcination at 600 °C for 4 h, producing the nanocapsules.

In a subsequent growth step, Ag nanocapsules (AgNCap) were formed within the previously prepared NCaps system[50]. The AuNPs confined inside the hollow cavity acted as nucleation sites for silver deposition. For this purpose, 0.5 mL of the NCaps suspension (1.25 mg $mL^{-1}$) was vigorously stirred while 1 mL of $AgNO_3$ solution (10 mM) was introduced. After 5 min, AA (10 µL, 10 mM) was added as a reducing agent, leading to the reduction of $Ag^+$ and a visible color change of the dispersion from red to brown. The reaction mixture was stirred for an additional 10 min, after which the product was isolated by centrifugation and washed with ethanol (3500 rpm, 10 min) to remove residual reagents, affording the AgNCaps. 10 µL of 1mM Raman reporters (4-FTP, 4-BrTP, 3,5 D($CF_3$)TP, BT and 4-MBA, respectively) were added into 1 mL AgNCaps to form SERS amplifiers.

The functionalized cover glass was prepared following the literature with slight modify[51]. The cover glass was sonicated with sequentially in acetone, ethanol, and ultrapure water for 10 min each, and then dried on a hot plate. The cleaned cover glass was immersed in a 10% (v/v) APTES ethanol solution at 70 °C for 2 h, followed by rinsing with ethanol three times. These 5 types of capsules were mixed and dropped on the APTES cover glass within 2 min, followed by gently flushing the sample area with ultrapure water.

**1 µm $SiO_2$@Ag microbeads in agarose**

1 mL aminated $SiO_2$ (1 mg/mL) was added into 100 mL of freshly prepared 80 nm Ag NPs solution as previously described and incubated overnight under gently rotation (20 rpm) by using tube rotator (*Fisher Scientific, Multi-Purpose Tube Rotator*). The resulting $SiO_2$@Ag microbeads were washed three times with ultrapure water and concentrated to a final volume of 1 mL.

Raman reporter stock solutions were diluted to 2 mM using ultrapure water. For labelling, 200 µL of $SiO_2$@Ag microbeads were mixed with 200 µL of 2 mM Raman reporter solution (4-MBA, 3-FTP, 4-MBT and BT) and incubated overnight. Before mixing, each type of microbead was washed three times with ultrapure water to remove the unbounding Raman reporter molecules. All the labelled microbeads were concentrated to 50 µL, mixed with 4% low-melting agarose, and then cooled at 4 °C immediately to form 3D agarose microbeads sample.

**Layered gelatine**

For 50 nm Ag NPs synthesis, a modified protocol was used[52]. Briefly, sodium citrate (2 mL, 34 mM), $AgNO_3$ (0.5 mL, 58.8 mM), and NaCl (0.4 mL, 20 mM) were added to 2.1 mL of ultrapure water, respectively, and stirred for 5 min. Then the mixture was poured into 95 mL of boiling water to which AA (160 µL, 100 mM) were added 1 minute before. After 30 min the solution was left to cool down. For each layer, 1 mL of NPs, as synthetized, and a volume of the Raman reported was added and then gently shacked. Employed volumes were 10, 7.1, and 12.5 uL of BT (1 mM), 4-MBA (10 mM), 4-FTP (10 mM), and 3-FTP (10 mM). Then 666 µL Gelatine (5%) was added, so a final (2%) concentration of gelatine was obtained. The solution was then left to set in the fridge overnight. The gelatine was left to partially dry in a fume hood for 3 h before the next layer was added. The final product was then completely dried, forming a single, merged material.

**Multiplex Bacteria@AgNP Substrates**

*E. coli* and *S. aureus* were used as representative bacteria to fabricate bacteria-based substrate using a modified protocol[53]. AgNPs were synthesized by using $NaBH_4$ as the reducing agent. Briefly, 13.2 mg of $NaBH_4$ was added to 99 mL of deionized water and stirred at 1200 rpm for 10 min. Subsequently, 6.6 mg of $AgNO_3$ was dissolved in 1 mL of deionized water and added to $NaBH_4$ solution under at a rate of 0.5 mL/s rate under continuous stirring at 1200 rpm. The reaction mixture was then stirred for an additional 2 h. After synthesis, the Ag NPs colloidal solution was centrifuged and washed 3 times. Then, 50 µL of 1mM NaCl and 50 µL of 1 mM $NaBH_4$ were sequentially added to 100 µL of the AgNPs colloidal solution. The resulting mixture was incubated in the dark for 4 h before further use.

Colonies of *E. coli* or *S. aureus* were picked from agar plates and inoculated into 5 mL of LB medium, followed by overnight incubation at 37 ℃ with shaking at 150 rpm. The bacterial suspensions were then centrifuged and washed 3 times. Then, 50 µL of 1 mM $NaBH_4$, 20 µL of bacterial suspension and 50 µL of 0.5 M NaCl were sequentially added to 100 µL of Ag NPs mixture solution and incubated for 10 min before further use. Next, bacteria@AgNPs suspensions were centrifuged at 2000 rpm, after which the supernatant was discarded, and the pellets were resuspended 100 µL of deionized water. Then, 100 µL of 2 mM of BT and 4MBA solutions were added in the *E. coli*@AgNPs and *S. aureus*@AgNPs suspensions, respectively, followed by incubation for 30 min. The labelled bacteria@AgNPs were centrifuged and washed 3 times. The labelled *E. coli*@AgNP and *S. aureus*@AgNP suspensions were then mixed to prepare the multiplex bacteria-based SERS substrate. Finally, the mixed bacterial sensor suspension was deposited onto poly-L-lysine-treated cover glass for subsequent measurements.

**Widefield SERS lock-in imaging**

Imaging parameters were adjusted according to the sample needs. The 1 µm $SiO_2$@Ag beads on PDMS film sample were imaged at a power density of 150 W/$cm^2$ with an acquisition time of 200 ms per frame. The 3 µm PS@Ag, AgNCaps, 3D gelatine, Bacteria@AgNP and 3D agarose samples were imaged at power densities of 1 kW/$cm^2$, 1 kW/$cm^2$, 1 kW/$cm^2$, 1 kW/$cm^2$ and 400 W/$cm^2$, respectively, each with an acquisition time of 100 ms per frame. For 3D imaging, a piezo (*PZ 200 OEM, Piezosystem Jena*) was integrated into the sample stage to enable Z-axis scanning. Upon completion of each 2D acquisition, the stage was stepped to the subsequent Z plane until the full volume was acquired. The z-step sizes were 1 µm and 5 µm for the agarose and gelatine samples, respectively, and the corresponding imaging volumes were 150.7×150.7×50 $\mu m^3$ and 175×175×60 $\mu m^3$.

**Data analysis**

The spectral data acquired from the spectrometer were wavelength-calibrated by converting pixel coordinates to wavelength values using a calibration lamp as a reference source. For each position of the interferometer arm, a corresponding spectrum was recorded, yielding a series of spectra whose number was determined by the total number of interferometer arm displacements. These spectra were subsequently subjected to FFT analysis to extract the time delay associated with each spectrum. The resulting time delays were then fitted using linear regression to establish a precise time-delay axis (Supplementary Information 1). Concurrently, SERS images were acquired using a camera, and the pixel intensities within the region of interest were spatially integrated for each image. The linearly fitted time delays were assigned as the x-axis, and the corresponding integrated pixel intensities from each SERS image were coupled to their respective time-delay values to construct the interferogram spectra, which served as the basis for subsequent FFT and lock-in analysis.

A Cellpose model was applied to a representative SERS image of each sample to segment and identify individual SERS amplifiers, microbeads, capsules, and nanoaggregates, and to generate corresponding masks[54]. Each mask was independently processed following the workflow described above, yielding its own interferogram spectra and SERS spectra. Classification of the SERS spectra was performed based on the Raman shift positions and intensity ratios of characteristic peaks. Upon classification, each mask was recoloured according to its assigned class and remapped onto the original-size image to produce a spatially resolved classification map. To identify the in-focus layer for agarose sample, Cellpose segmentation was performed on each z-stack to generate masks. Each mask was subsequently tracked across the following 5 z-layers, yielding 6 candidate layers in total. Detections were matched across candidate layers based on lateral displacement, and the layer yielding the highest pixel intensity was assigned as the in-focus layer, as the diffraction pattern of a spherical bead is most compact at focus.

All the spectra of masks were classified and recoloured in the 3D spatial scatter plot (Supplementary Information 4).

## Data Availability,

The data and code that support the findings of this study are available from the corresponding author upon reasonable request.

## References,

# Supplementary Information:

## Sub-Nyquist time-domain surface-enhanced Raman mapping

Ting Wang[1,*], I. Brian Becerril-Castro[2,*], Ana Sousa-Castillo[3], Miguel A. Correa-Duarte[3,4,5], Ramón A. Alvarez-Puebla[2,6], Matz Liebel[1,a]

[1] Department of Physics and Astronomy, Vrije Universiteit Amsterdam, De Boelelaan 1100, Amsterdam 1081 HZ, The Netherlands

[2] Department of Physical and Inorganic Chemistry, Universitat Rovira i Virgili, Carrer Marcel·lí Domingo 2-4-6, 43007, Tarragona, Spain

[3] CINBIO, Universidade de Vigo, Vigo 36310, Spain

[4] Southern Galicia Institute of Health Research (IISGS), Vigo 36310, Spain

[5] Biomedical Research Networking Center for Mental Health (CIBERSAM), Vigo 36310, Spain

[6] ICREA, Passeig Lluís Companys 23, 08010, Barcelona, Spain

[*] Both authors contributed equally to this work.

[a] email: m.liebel@vu.nl

### Supplementary Information 1: Time-delay measurements and SERS spectra retrieval

A custom spectrometer (Methods, Main Manuscript) was calibrated based on Ne/Ar reference lines and then used to detect the time-delay modulated super luminescent diode spectrum after it passed the interferometer. At each interferometer arm position, a spectrum was recorded, in parallel to the widefield images of the microscope imaging camera. The time-delay retrieval is based on two consecutive operations. We first retrieve a coarse mean step-size estimate which we then use as a guess for a direct spectral fit that extracts the precise time-delay from the modulated spectra.

To implement the former, we selected a fraction of the data, starting shortly after time-zero, e.g. within 50-200 fs to avoid difficulties associated with accurately determining positions close to time-zero where the symmetric temporal components at positive and negative delays are close to each other. We first subtracted the mean over all spectra from each individual spectrum and then interpolated all spectra onto a linear frequency axis. We then zero padded to 16384 points followed by Fast Fourier transformation. A time-delay guess was obtained as the temporal location of maximum amplitude of the resulting time-domain representation. Only positive time-delays were considered to avoid fit-ambiguities, as mentioned above. Using these results, we extracted mean step-size and time-zero estimates via a linear fit.

Using these parameters we computed time-delay guesses for all measurement points. These guesses were then used to obtain refined time-delays, $\Delta t$, by directly fitting each individual spectral measurement, $I_{spec}$, as:

$$I_{spec}(\Delta t,\omega) = Re\{a^2A(\omega)^2 + b^2A(\omega)^2 + abA(\omega)^2 exp[2\pi i\omega\Delta t] + abA(\omega)^2[-2\pi i\omega\Delta t]\}$$

With $\omega$ being the frequency, $\Delta t$ the time-delay of interest and $A(\omega)$ the spectral amplitude obtained as the square root of the median over all spectral measurements. The constants $a,b$ account for the fact that the amplitudes obtained via the two nominally identical interferometer arms might be slightly different.

### Supplementary Information 2: Retrieved vs. simulated spectra

Supplementary Figure 1 compares the experimentally retrieved spectra shown in (Figure 2d, Main Manuscript) to simulated spectra thus providing a rational understanding of the aliasing-induced spectral shifts.

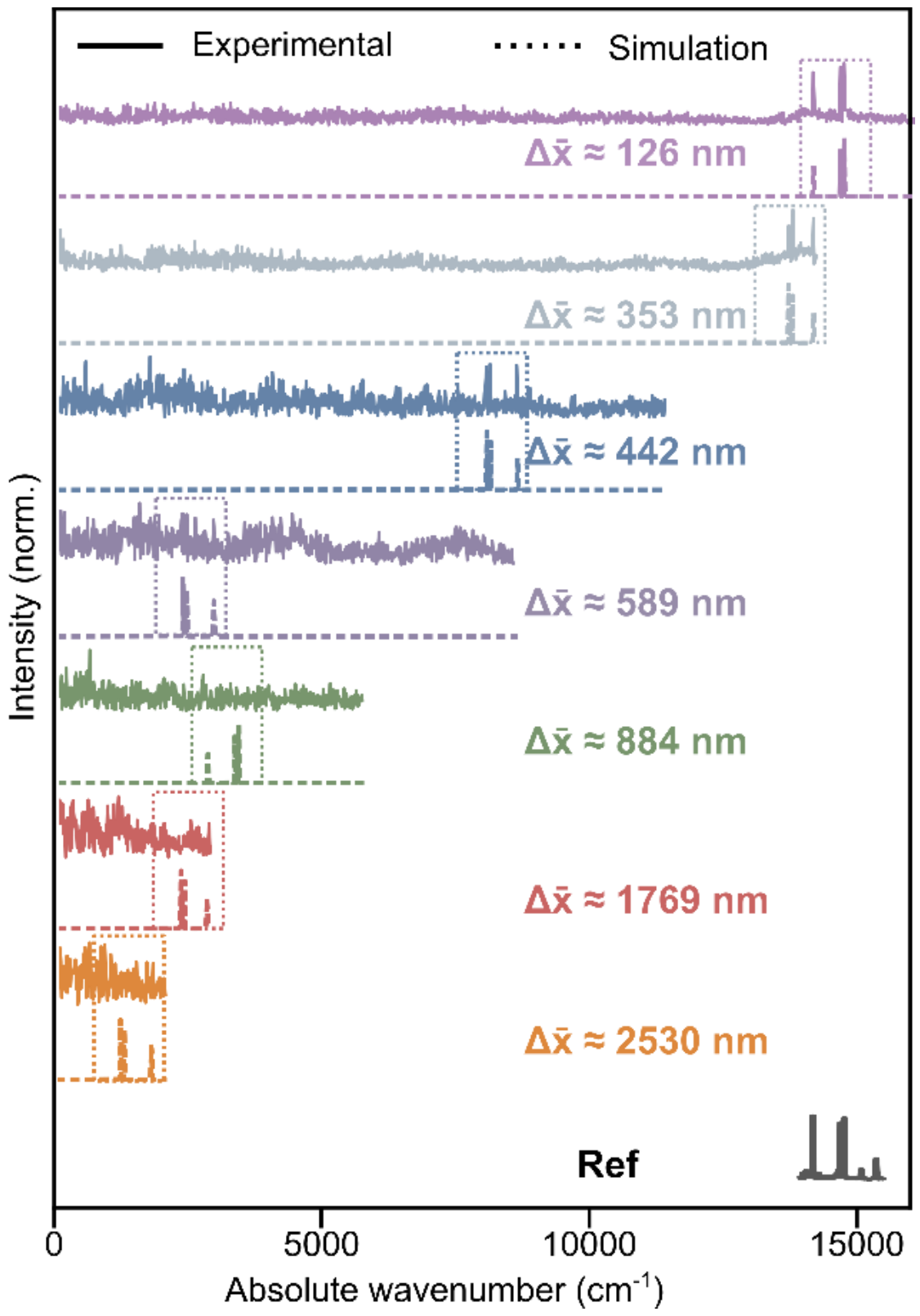

**Supplementary Figure 1,** Comparison of the SERS spectra obtained via a Fast Fourier transformation of experimental (solid lines) and simulated (dashed lines) data.

## Supplementary Information 3: Detection parameter optimization

To optimize the sampling conditions, as outlined in the main text, we relied on a simple simulation using typical frequencies as obtained by a Ne/Ar calibration lamp (converted to characteristic peaks in Raman shifts at 695, 770, 1250, and 1450 $cm^{-1}$). A bandwidth of 15 $cm^{-1}$ was assumed in the simulation. We systematically varied step sizes between 1 μm and 3 μm assuming a total of 2048 steps (Supplementary Figure 2a). To maximize acquisition speed while avoiding aliasing, we identified a step size range of 2.5~2.7 μm, yielding visually "clean" spectra. Supplementary Figure 2b, c compare simulation results and corresponding experimental measurements. Beyond appearance, spectral resolution was also considered as an optimization parameter. The number of steps was systematically varied from 128 to 1024, and both simulated and measured results again demonstrate high consistency. However, since the calibration lamp emits atomic lines, the peaks become increasingly narrow as the step number increases (Supplementary Figure 2d, e), which should be taken into account when selecting the appropriate step number given that room-temperature Raman spectra typically exhibit linewidth in the 10-20 $cm^{-1}$ range. Thus, to balance spectral resolution, observation window, and acquisition speed, the optimal measurement parameters were determined to be a step size of 2.6 μm with 350 steps.

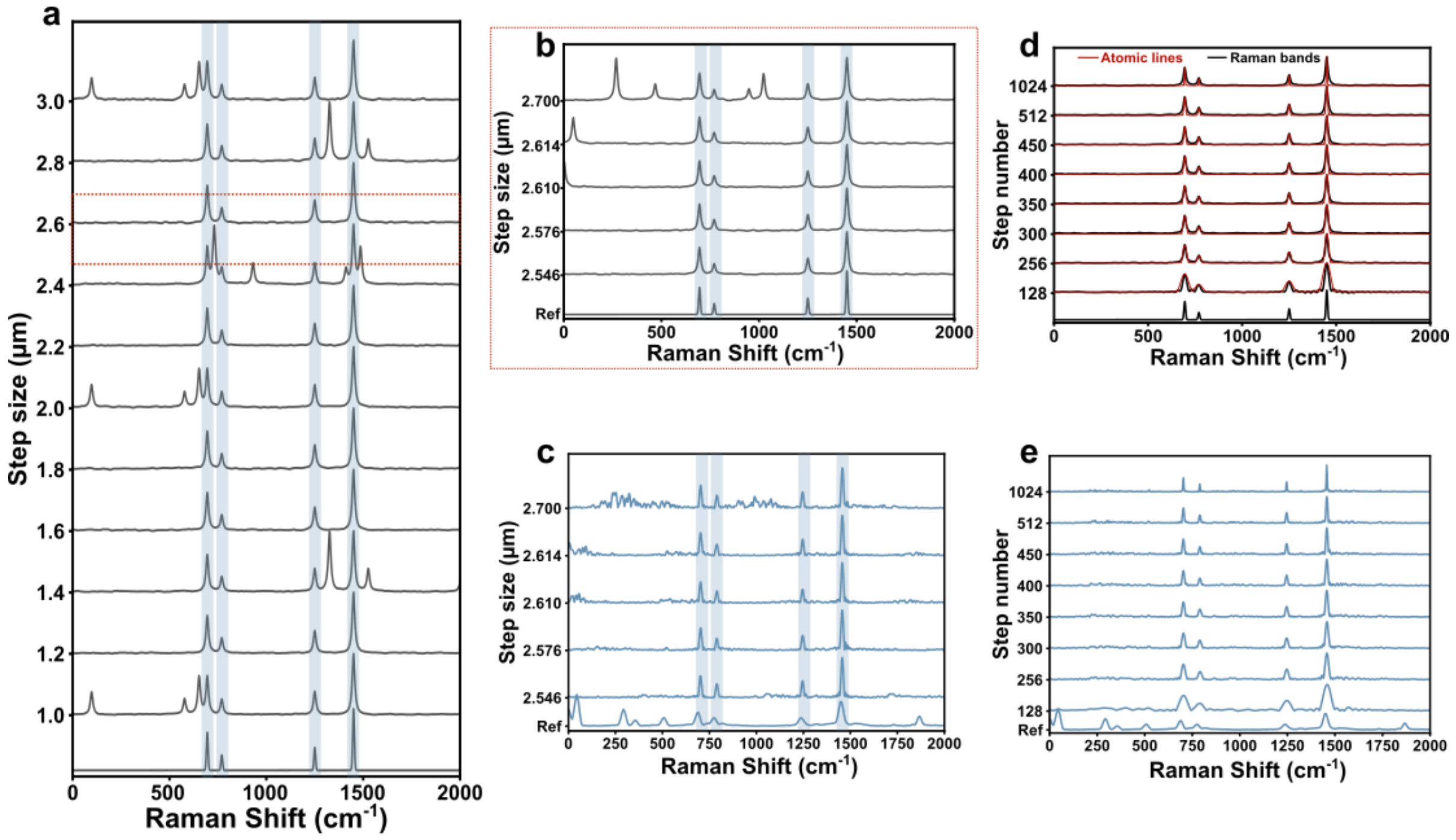

**Supplementary Figure 2,** Optimization of interferometric scanning parameters through simulation and experimental validation. (**a**) Simulated Raman spectra obtained at different step sizes (1-3 μm) using Raman shifts corresponding to Ne/Ar atomic lines. Blue bars highlight the shifts used in the simulation, the other bands are aliasing-induced artefacts. (**b**) Refined simulations for step sizes between 2.5 μm and 2.7 μm to identify the optimal range for undistorted spectral retrieval. (**c**) Experimental validation using a calibration lamp under the same conditions as in (b). (**d**) Simulated Raman spectra (black line) and atomic lines (red line) of calibration lamp for step-numbers ranging from 128-1024. (**e**) Experimental validation of (d) using the calibration lamp combined with a step size of 2.6 μm. Raman simulations were conducted assuming a bandwidth of 15 cm$^{-1}$.

## Supplementary Information 4: 3D localization and focused layer determination of SERS amplifiers

To determine the in-focus z-position of each SERS amplifier in the 3D agarose samples (Figure 4, Main Manuscript), the Cellpose-derived microbead probability maps were thresholded at a logit value of -3 to generate binary masks for each z-layer. Connected component analysis was subsequently applied to each binary mask to identify individual objects and extract their centroids and cross-sectional areas. Objects with fewer than 20 pixels were discarded as noise. To associate detections across z-layers, each object identified in a given layer was matched to detections in the subsequent five layers based on a maximum lateral displacement threshold of 2 pixels. Among all candidate layers for a given object, the layer yielding the highest pixel intensity was designated as the in-focus layer, as diffraction-limited imaging of spherical beads produces the most compact intensity distribution at the focal plane. The 3D coordinates of each bead were determined from the centroid position at the in-focus layer.

Based on observations from the agarose sample, SERS-active microbeads typically appear across 5 consecutive z-layers. For the gelatine sample, which has a z-step of 5 µm between individual layers, this characteristic reduces the likelihood of out-of-focus contributions. For sake of simplicity we, therefore, applied Cellpose independently to each z-layer to identify SERS-active microbeads. Supplementary Figure 3 shows the typical through-focus evolution of a set of microbeads.

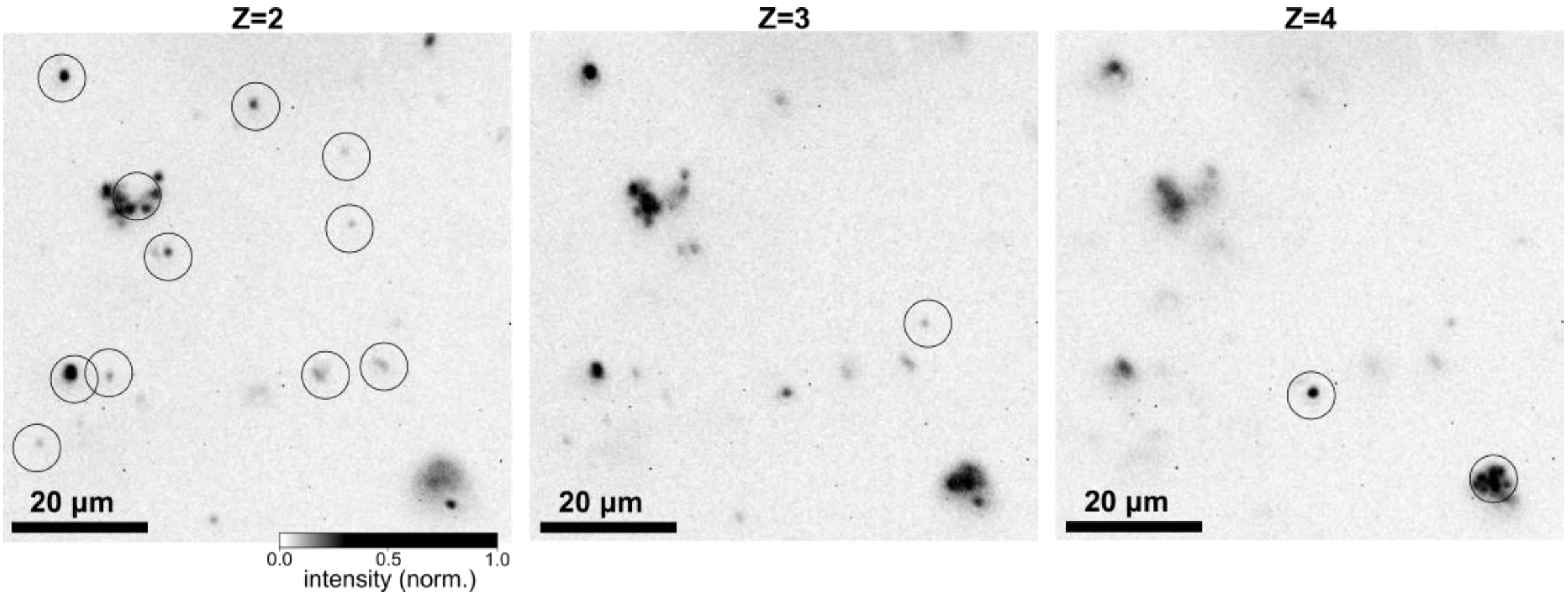

**Supplementary Figure 3,** Enlarge stack of the 2,3,4 stack and corresponding SERS amplifiers which attributed to this stack.

### Supplementary Information 5: Spectral Quality Improvement and Calibration

SERS lock-in retrieval, as well as most Fourier-transform experiments, relies on sampling "half" an interferogram ranging from around time zero to positive time delays. This sampling strategy is justified given that the data is necessary symmetric but results in spectral broadening and non-ideal line shapes. To improve the spectral quality, the interferograms were thus mirrored prior to lock-in retrieval. Importantly, mirroring shifts the time delay axis, which must be adjusted accordingly. The properly mirrored interferogram yields improved spectral quality in terms of both resolution and intensity. All spectra used for classification were obtained from mirrored interferograms. Supplementary Figure 4 highlights the problem of non-mirrored retrieval and the improvements achieved. To implement this strategy, precise knowledge of time-zero is necessary as destructive interference between the now symmetric signal can result in severe line shape distortions and loss-of-amplitude, as shown in Supplementary Figure 4. An alternative strategy for spectral recovery in the context of the time-zero problem which impacts the line shape if not accounted for is to rely on phase-recovery by analysing the central part of the interferogram around time-zero[1] .

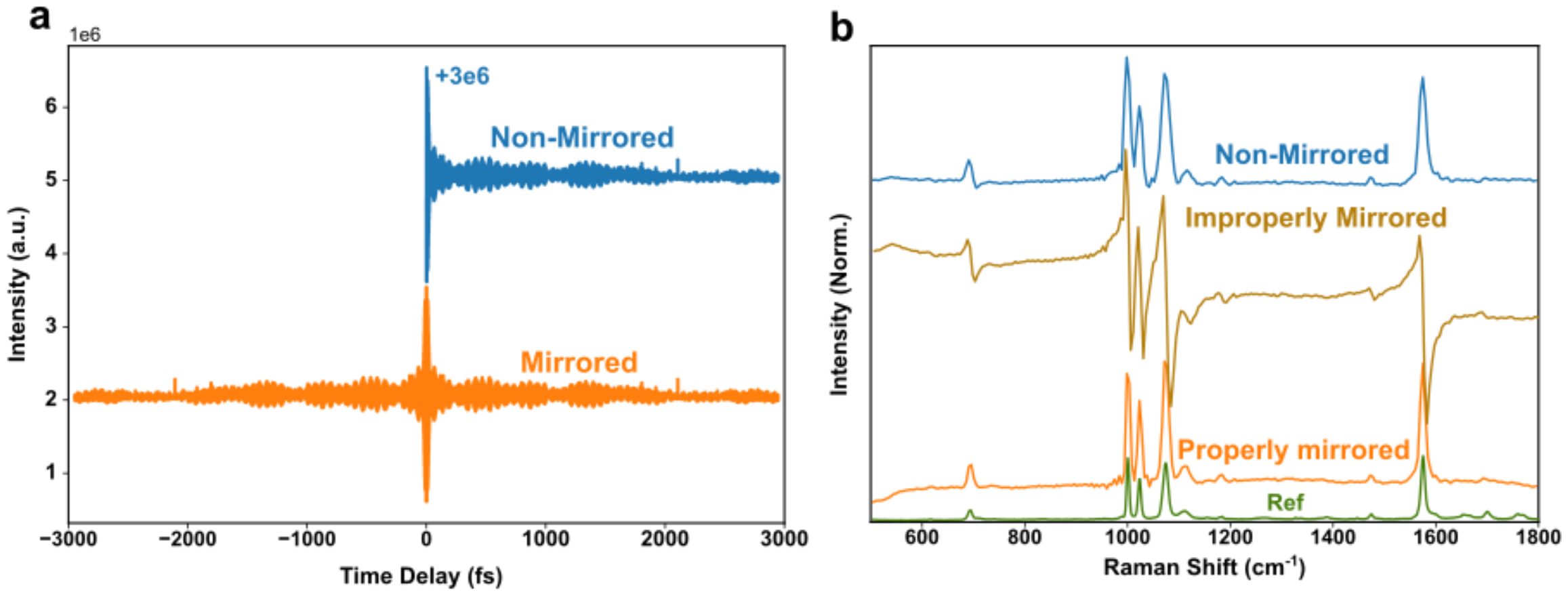

**Supplementary Figure 4,** (a) The original and typical mirrored interferograms. (b) SERS spectra after lock-in retrieval from the original interferogram, the properly mirrored interferogram (correct time-axis), and an improperly mirrored interferogram (incorrect time-axis). The spectra are offset for clarity.

We noted that field curvature made the experimentally measured Raman shifts detector position dependent, e.g. nominally identical particles would show slightly shifted frequencies depending on where they were located within the image. To correct for this aspect, a calibration lamp was used to determine the frequency shifts and then correct for them. Towards this goal, spectra were recorded at every 72nd pixel on the camera, with the frequency at the centre pixel taken as the reference standard. The peak centred at approximately 1455 $cm^{-1}$ was selected as the calibration peak to determine the shift at each pixel position. The peak shifts across all sampled pixels were then interpolated using a linear interpolator to generate a continuous shift field over the full field of view. For the actual sample, the mean peak shift for each SERS amplifier was calculated based on the spatial extent of its corresponding mask, and this mean shift was subsequently applied to correct the Raman shift of the associated SERS spectra. However, the uncorrected SERS spectra were used for classification in this work, as the molecular SERS fingerprint features, such as the number and relative ratios of peaks, provide sufficient discriminative information for accurate classification.

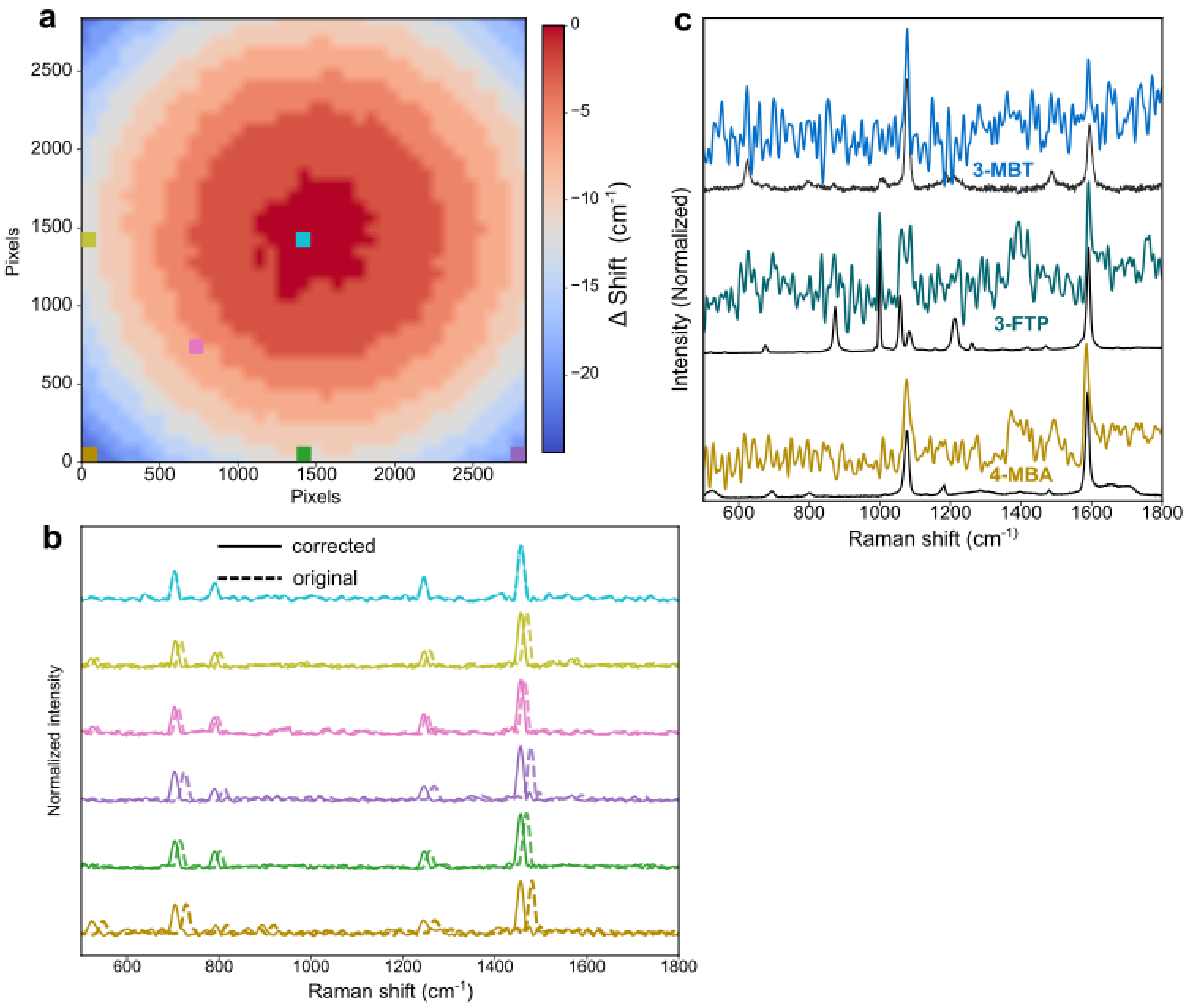

**Supplementary Figure 5,** a) Spatial distribution of Raman shifts deviation (ΔShift, $cm^{-1}$) across the field-of-view. The cyan square marks the reference centre pixel (ΔShift = 0) and coloured squares indicate representative pixel positions. (b) Corresponding Raman spectra at the selected pixel positions. (c) Representative corrected SERS spectra (coloured) from a highly multiplexed sample containing multiple analytes (4-MBT, 3-FTP, and 4-MBA) on 1 µm $SiO_2$@Ag microbeads, alongside their standard SERS spectra (black).